\documentclass[usenatbib]{mn2e}
\usepackage{fixltx2e}
\usepackage{mathptmx}
\usepackage{color}
\usepackage{graphics,graphicx,amssymb,amsmath,fixltx2e,natbib}

\newcommand{\apj}{ApJ}
\newcommand{\mnras}{MNRAS}

\newcommand{\araa}{ARA\&A}                
\newcommand{\aap}{A\&A}                   
\newcommand{\apjs}{ApJS}                  
\newcommand{\apjl}{ApJ}                   
\newcommand{\pasj}{PASJ}
\newcommand{\gca}{Geochim. Cosmochim. Acta}

\def\gtrsim{\lower 2pt \hbox{$\, \buildrel {\scriptstyle >}\over
{\scriptstyle \sim}\,$}}
\def\lesssim{\lower 2pt \hbox{$\, \buildrel {\scriptstyle <}\over
{\scriptstyle \sim}\,$}}

\def\ROSAT{{\sl ROSAT}}

\def\hst{{\sl HST}}

\def\chandra{{\sl Chandra}}

\def\HST{{\sl HST}}
\def\hst{{\sl HST}}

\newcommand{\Lya}{Ly-$\alpha$}
\newcommand{\Lyb}{Ly-$\beta$}

\def\xs{{HS 1700+6416}}

\begin{document}

\title{X-ray mapping the outer regions of galaxy clusters at z= 0.23 and 0.45}
\author[]{Q. Daniel Wang$^{1,2}$\thanks{E-mail:wqd@astro.umass.edu} and Stephen Walker$^{2}$\\ 
$^{1}$Department of Astronomy, University of Massachusetts,  Amherst, MA 01003, USA\\
$^{2}$ Institute of Astronomy, University of Cambridge, Madingley Road, Cambridge CB3 0HA, UK.\\
}



\maketitle

\label{firstpage}

\begin{abstract}
The thermal, chemical, and kinematic properties of the
potentially multi-phase circum/inter-galactic medium at the virial radii
of galaxy clusters remain largely uncertain.
We present an X-ray study of Abell 2246 and GMBCG J255.34805+64.23661
($z=0.23$ and 0.45), two foreground clusters of the
UV-bright QSO HS 1700+6416, based on 240 ks {\sl Chandra}/ACIS-I
observations. We detect enhanced diffuse X-ray emission to the
projected distances beyond $r_{200}$ radii of these two clusters.  
The large-scale X-ray emission is consistent with being azimuthally
symmetric at the projected radii of the QSO (0.36 and 0.8 times the 
radii of the two clusters). Assuming a spherical symmetry, 
we obtain the de-projected temperature and density profiles
of the X-ray-emitting gas. Excluding the cool cores that are
detected, we find that the mean temperature of the hot gas is 
$\sim 4.0$~keV for Abell 2246 and 5.5~keV for GMBCG J255.34805+64.23661, 
although there are indications for temperature drop at large radii. From these
results, we can estimate the density and pressure distributions of the
hot gas along the QSO sightline. We further infer the radial entropy
profile of Abell 2246 and compare it with the one expected from purely gravitational
hierarchical structure formation. This comparison shows that the ICM
in the outer region of the clusters 
is likely in a clumpy and multi-phased state.
These results, together with the upcoming 
{\sl HST}/COS observations of the QSO sightline, will enable a comprehensive 
investigation of the multi-phase medium associated with the clusters. 
\end{abstract}


\section{Introduction}
A major recent advance in the study of nearby galaxy clusters is the detection
of apparently diffuse X-ray emission from their outer regions at radius 
$r \sim r_{500}-r_{200}$, where $r_{200}$ (or $r_{500}$)  is the
radius within which the mean mass density is 200 (or  500) times the
critical density of the universe at the redshift of a cluster~\citep[e.g.,][]{George2009,Hoshino2010,Simionescu2011,Eckert2012,Walker2013}. 
In particular, there is now strong 
evidence for the inhomogeneous distribution of the intracluster medium 
(ICM) in these regions.
This inhomogeneous and hence potentially multi-phase nature of the 
ICM is expected from simulations of the structure 
formation of the universe, which show a complicated shock heating/cooling history of the ICM in 
regions near the virial shock, typically located at $r \sim r_{200}$
of a cluster  \citep[e.g.,][]{Roncarelli2006,Molnar2009}. 
The clustering environment can also strongly affect the circumgalactic medium (CGM) of individual galaxies, via such processes as ram-pressure
stripping and pressure compression \citep[e.g.,][]{Yoon2013,Lu2011}. 
Understanding these phenomena and physical processes is clearly important, not 
only for determining the content and state of the baryon matter in the outer regions of clusters, 
but for their utility as cosmology probes as well (e.g., via the 
observations of the Sunyaev-Zel'dovich effect; \citealt{Carlstrom2002}). 

However, it has been challenging for existing studies to cover
the large angular extents of nearby clusters and to calibrate out 
various systematics, which are important for mapping out the low
surface intensity of the X-ray emission.
Indeed, much of the advance has been made, largely because of the extensive 
observations made with {\sl Suzaku}/XIS, which has a low instrumental 
background. But, the studies are fundamentally limited by various uncertainties,
such as the cosmic variance (in the surface density of discrete and extended 
X-ray sources due to the structured nature of the universe; \citealt{Hickox2006}), 
the intensity and spectral variation of the Galactic soft X-ray foreground 
across large sky areas and the time-dependent 
contributions from solar wind charge exchanges in the
earth's magnetosphere and/or the heliosphere~\citep{Koutroumpa2006}.

Complementary investigations can be conducted with deep \chandra\ observations
of distant clusters ($z \gtrsim 0.2$). 
The superb spatial resolution and sensitivity of such observations allow
for the mapping of the X-ray emission from clusters
with minimal confusion from spurious sources and little differential 
foreground variation. In a single observation 
(especially with the ACIS-I at the aim point), both a cluster and its local 
sky background can be covered. In addition to these advantages, such 
investigations are, of course, essential to the understanding of the
galaxy cluster building process across the cosmic time. 
\citet{Bonamente2013} have reported the first \chandra\ detection of 
diffuse X-ray emission out to the virial radius of the cluster Abell 1835 
at $z = 0.253$. A sharp drop in temperature is detected at large radii, 
indicating that that the ICM there is significantly clumped and/or contains multiple phases.

\begin{table*}
\begin{center}
\begin{minipage}[b]{4in}
\caption{{\sl Chandra} ACIS-I Observations}
\begin{tabular}{@{}lccccc}
\hline\hline
OBS\# &R.A. (J2000) &Dec. (J2000) & Exposure &Roll Angle &OBS Date\\
 &(h~~m~~s) &($^\circ~~~^{\prime}~~~^{\prime\prime}$)&(s) &($^\circ$) &(yyyy-mm-dd)\\
\hline
547      &17 01 46.14 & +64 12 04.3 &  46967 &   322.4 &2000-10-31 \\
8032     &17 00 55.04 & +64 11 26.7 &  30996 &   338.0 &2007-11-12 \\
8033     &17 00 55.05 & +64 11 26.7 &  29707 &   338.0 &2007-11-20 \\
9756     &17 00 55.05 & +64 11 26.7 &  31234 &   338.0 &2007-11-14 \\
9757     &17 00 55.03 & +64 11 26.7 &  20527 &   338.0 &2007-11-13 \\
9758     &17 00 55.05 & +64 11 26.6 &  23114 &   338.0 &2007-11-16 \\
9759     &17 00 55.03 & +64 11 26.7 &  31187 &   338.0 &2007-11-17 \\
9760     &17 00 55.06 & +64 11 26.7 &  16694 &   338.0 &2007-11-19 \\
9767     &17 00 55.02 & +64 11 27.0 &   9039 &   338.0 &2007-11-21 \\
\hline
\end{tabular}

\medskip
The exposure represents the livetime (dead time corrected) of the
cleaned data.
\label{t:obs}
\end{minipage}
\end{center}
\end{table*}

\begin{table*}
\begin{center}
\begin{minipage}[t]{4in}
\caption{Key parameters of the clusters}
\begin{tabular}{@{}lcc}
\hline\hline
Cluster  & A & B \\
\hline
Name & Abell 2246 & GMBCG J255.34805+64.23661\\
Redshift & 0.23 & 0.45\\
Center position  &$17^h00^m41.8^s, +64^\circ12^{\prime}59^{\prime\prime}$&$17^h01^m23.5^s, +64^\circ14^{\prime}12^{\prime\prime}$\\
$L_{0.5-8\rm~keV} ({\rm erg~s^{-1}})$ & 1.9$\times$10$^{44}$ &5.4$\times$10$^{44}$ \\
Temperature (K) & $4.0^{+0.2}_{-0.3}$ & $5.5^{+0.4}_{-0.4}$\\
$M_{200} (10^{14} M_\odot)$ & $3.3^{+0.6}_{-0.4}$  &$5.0^{+0.8}_{-0.7}$\\
$r_{200}$ ($^\prime$/Mpc) & $6.1^{+0.5}_{-0.3}/1.3^{+0.1}_{-0.1}$& $4.0^{+0.2}_{-0.2}/1.4^{+0.1}_{-0.1}$\\
\\
$I_0 ({\rm cts~s^{-1}~arcmin^{-2}})$ & 0.21&0.42\\
$r_c/r_{200}$ &0.026 &0.016\\
$\beta$ & 0.51&0.45\\
$\chi^2/d.o.f.$ & 79/36 & 46/36\\
\\
QSO Impact parameter & 2.2$^\prime$ & 3.2$^\prime$\\ 
$N_e (10^{20}{\rm~cm^{-2}})$&6.5&4.0\\
\hline
\end{tabular}
Note: For each of the two cluster, the name and redshift are from NED,
position from the present
\chandra\ source detection, X-ray luminosity and temperature from the
best-fit to the spectrum extracted from the region roughly within 
the projected $r_{500}$, the cluster mass ($M_{200}$) and $r_{200}$ from their scaling
relations with the temperature~\citep{Arnaud2005},
$\beta$-model parameters (central count rate $I_0$, projected core radius $r_c$, and $\beta$; see Eq.~\ref{e:beta}) 
from the best-fit to the 0.5-2~keV intensity profile,
electron column density $N_e$ at the projected radius of \xs. 
\label{t:para}
\end{minipage}
\end{center}
 \end{table*}

\begin{figure*}
\unitlength1.0cm
\centerline{
\includegraphics[width=0.95\linewidth,angle=0]{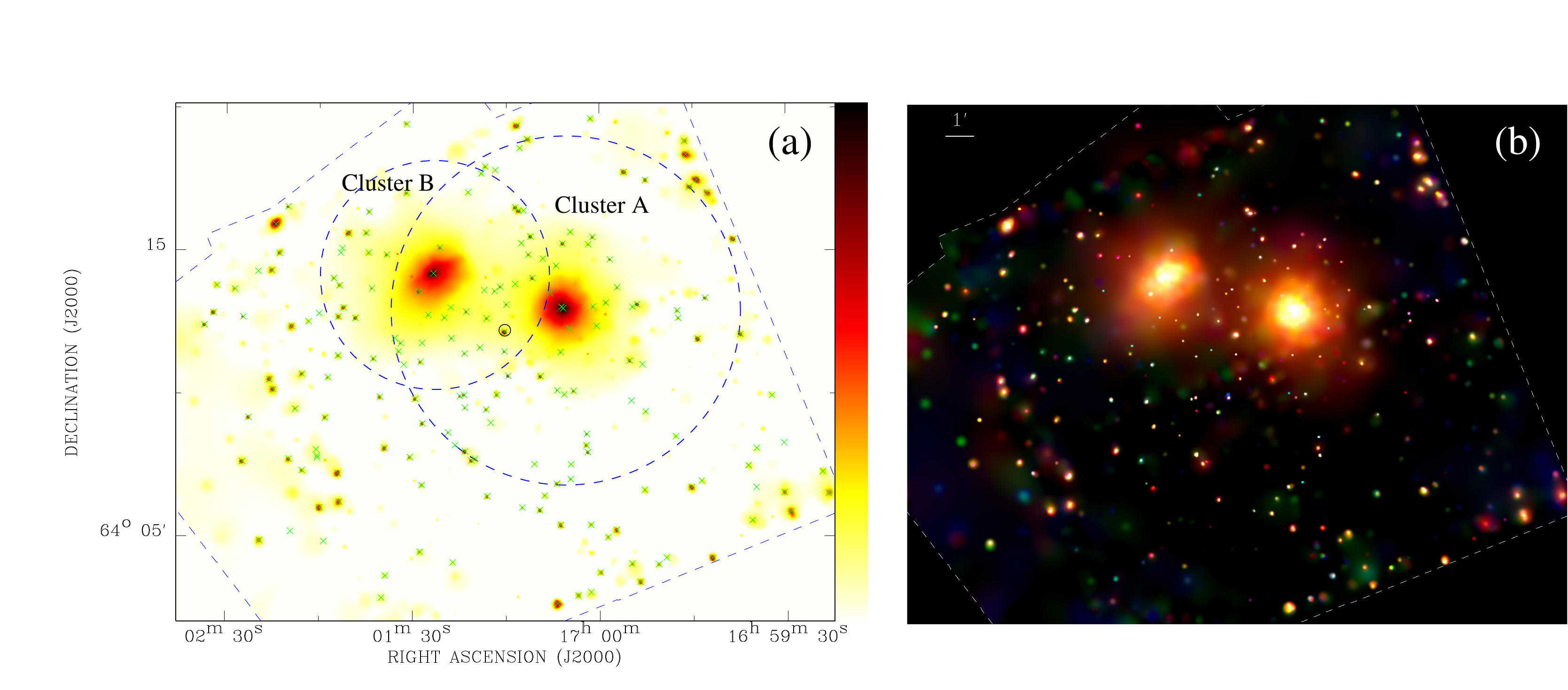}
}
\caption{X-ray intensity images of our studied field in various ACIS-I
  bands. (a) Image in the B (0.5-8 keV) band. The positions of the detected sources are marked
with {\sl crosses}, whereas \xs\ is highlighted by a small
circle. (b) Tri-color montage of X-ray intensities in the same field: (red) 0.5-2 keV, (green) 
2-4 keV, and (blue) 4-8 keV. The dashed contour outlines the boundaries of the
  field covered by the merged ACIS-I data, while the large circles mark the $r_{200}$ radii of the clusters.
}
\label{fig:field_total_sou}
\end{figure*}

Here we present a study of two intermediate redshift clusters, Abell 2246
and GMBCG J255.34805+64.23661 (Clusters A and B hereafter; Table~\ref{t:obs})
based
on nine \chandra\ observations of a total 240 ks exposure. 
The X-ray emission from these two clusters (Fig.~\ref{fig:field_total_sou}), 
are serendipitously discovered by {\sl ROSAT} observations of 
QSO HS 1700+6416 ($z=2.72$; ~\citealt{Reimers1997,Just2007,Lanzuisi2012}). 
This QSO is 2.2$^\prime$ and
3.2$^\prime$ off the center of Cluster A and B, respectively. These QSO/cluster alignments, well
within the projected virial radii of the clusters 
(Fig.~\ref{fig:field_total_sou}; see \S~\ref{s:res})
allow for various absorption line studies, especially at far-UV with upcoming
\hst/COS observations, which will be sensitive to OVI doublet (1032 and 1037 \AA),
HI \Lya\ and \Lyb, as well as many other transitions. These lines can 
offer unique information about the thermal, kinematic, and/or
chemical properties of cold and warm gases. The combination of
such a multi-wavelength studies will enable us to probe all phases of
the ICM, as well as the CGM
associated with individual galaxies, at multiple impact radii. 
The early \chandra\ observation, OBS\# 547 (Table~\ref{t:para}), 
has been included in various X-ray surveys
of clusters (e.g., ~\citealt{Maughan2008,Ettori2009}). 
The other eight observations have been used to investigate 
a $z = 2.30$ protocluster in the field~\citep{Digby-North2010}. While the 
diffuse X-ray emission from this protocluster is not detected, two
discrete X-ray sources are identified as the counterparts of emission
line AGNs in this protocluster. 

The focus of the present paper
is on Clusters A and B. We use the nine observations together
to characterize the properties of the hot ICM of the clusters
along the QSO sightline, as well as their global X-ray structures.
We describe the nine observations
and our reduction of the data in \S~2, present our data analysis 
in \S~3, and infer the radial properties of the ICM in \S~4. In S~5, we
summarize our results  and
discuss their implications and expected synergy
with the UV absorption line spectroscopy.
We use the standard 
$\Lambda$CDM cosmology and present the error bars of our measurements 
at the 68\% statistical confidence level, unless noted otherwise.

\section{Observations and Data Reduction}\label{s:obs}

The \chandra\ data used in the present work were taken in nine
separate observations (Table~\ref{t:obs}). We reprocess the
data with the latest version of the \chandra\ Interactive 
Analysis of Observations (CIAO; version 4.5). 
We remove time intervals of significant background flares with count rates
deviating more than 3$\sigma$ or a factor of $\gtrsim 1.2$ from the mean rate 
of individual observations, 
using Maxim Markevitch's light-curve cleaning routine \textsc{LC$\_$CLEAN}. 
This cleaning, together with a correction for the dead time of the 
observations, resulted in a total 239.5 ks exposure for subsequent
analysis. 

OBS 547 was taken in 2000, about seven years earlier than the
rest of the observations, which were all within a period of 10 days and with the
essentially same pointing positions and roll angles. Both the instrument
response and the non-X-ray background level
were significantly different between these two time periods.  
Thus OBS 547, also with a pointing offset of $\sim 5.6^\prime$ from the rest, is dealt
with separately, in both imaging and spectral analyses. This offset between
the two data sets allows us to test the sensitivity of our various
measurements on the detector coverage and source detection sensitivity.
We repeat the measurements such as radial X-ray intensity
distributions by using the two data sets separately to test the
consistency. Based on such tests, for example, we decide to
avoid the use of the regions with effective exposures $\lesssim 20$ ks in
our quantitative measurements to minimize the effects of low counting
statistics and the incomplete mapping of the field, mostly in its outermost edges. 

We estimate non-X-ray background contributions, using the ACIS stowed 
background database (Group ``D'' for OBS\# 547 and ``E'' for the rest). 
The charge-transfer inefficiency and gain 
corrections of the database are all matched to those of the present 
observations. Both the
data and background events are cleaned for the VFaint data mode.
The stowed background level in each observation is further normalized according to 
the rates in the 9-12 keV band, where events are almost completely due to the non-X-ray 
background.

We construct effective exposure maps in the four bands: 0.5-1 (S1),
1-2 (S2), 2-4 (H1), and 
4-8 (H2) keV. The construction of these broad-band exposure maps assumes a power 
law spectrum of photon index 1.7 and accounts for the
telescope vignetting and bad pixels as well as the 
quantum efficiency variation of the instrument, 
including the time-dependent sensitivity degradation, which is 
particularly important at energies $\lesssim 1$ keV. 

\section{Analysis and Results}\label{s:res}

Fig.~\ref{fig:field_total_sou}a presents an overview of our interested
field in the ACIS-I B (0.5-8 keV) band. This presented image is the sum of two intensity
images constructed  in the S (0.5-2 keV) and H (2-8 keV) bands and 
adaptively smoothed with the CIAO routine \textsc{csmooth}. The smoothing scales are calculated separately in the two bands and with the S/N ratio $\sim 3$. However, our quantitative analysis of the clusters is always based on the
unsmoothed count, background, and exposure maps. We describe our
detection of discrete X-ray sources in Appendix~\ref{a:app}. These sources
are marked in Fig.~\ref{fig:field_total_sou}a.
 Depending on the spectrum of a source, it may appear significant in
the S or H band, sometimes not in the B band. This band-dependence of the
X-ray emission can be better appreciated in the composite image
presented in Fig.~\ref{fig:field_total_sou}b. 

\subsection{Imaging analysis of the diffuse X-ray emission}

\begin{figure*}
\unitlength1.0cm
\centerline{
\includegraphics[width=0.95\linewidth,angle=0]{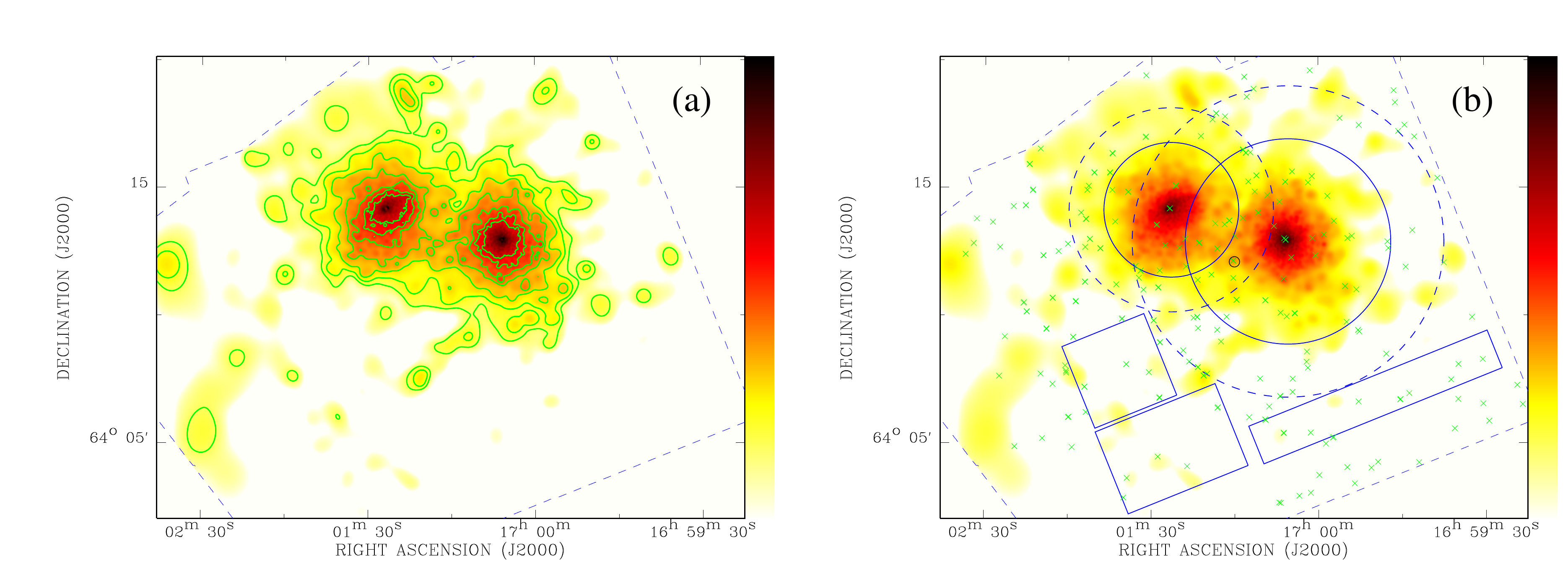}
}
\caption{Diffuse 0.5-2 keV emission intensity map of the field: (a)
  The contours (in units of $10^{-4} {\rm~cts~s^{-1}~arcmin^{-2}}$) are at
  4.5, 6, 9, 15, 27,   51, and   99; the lowest level is about
  3$\sigma$ above the local sky background of an intensity $\sim 3$; (b) The same image as in (a), but
 with the outlines of the spectral extraction
  regions for the two on-cluster regions (solid circles) and three
  off-cluster background regions (solid boxes). The rest is the same as
  in Fig.~\ref{fig:field_total_sou}.
}
\label{fig:field_dif}
\end{figure*}

\begin{figure*}
\unitlength1.0cm
\centerline{
\includegraphics[width=0.95\linewidth,angle=90]{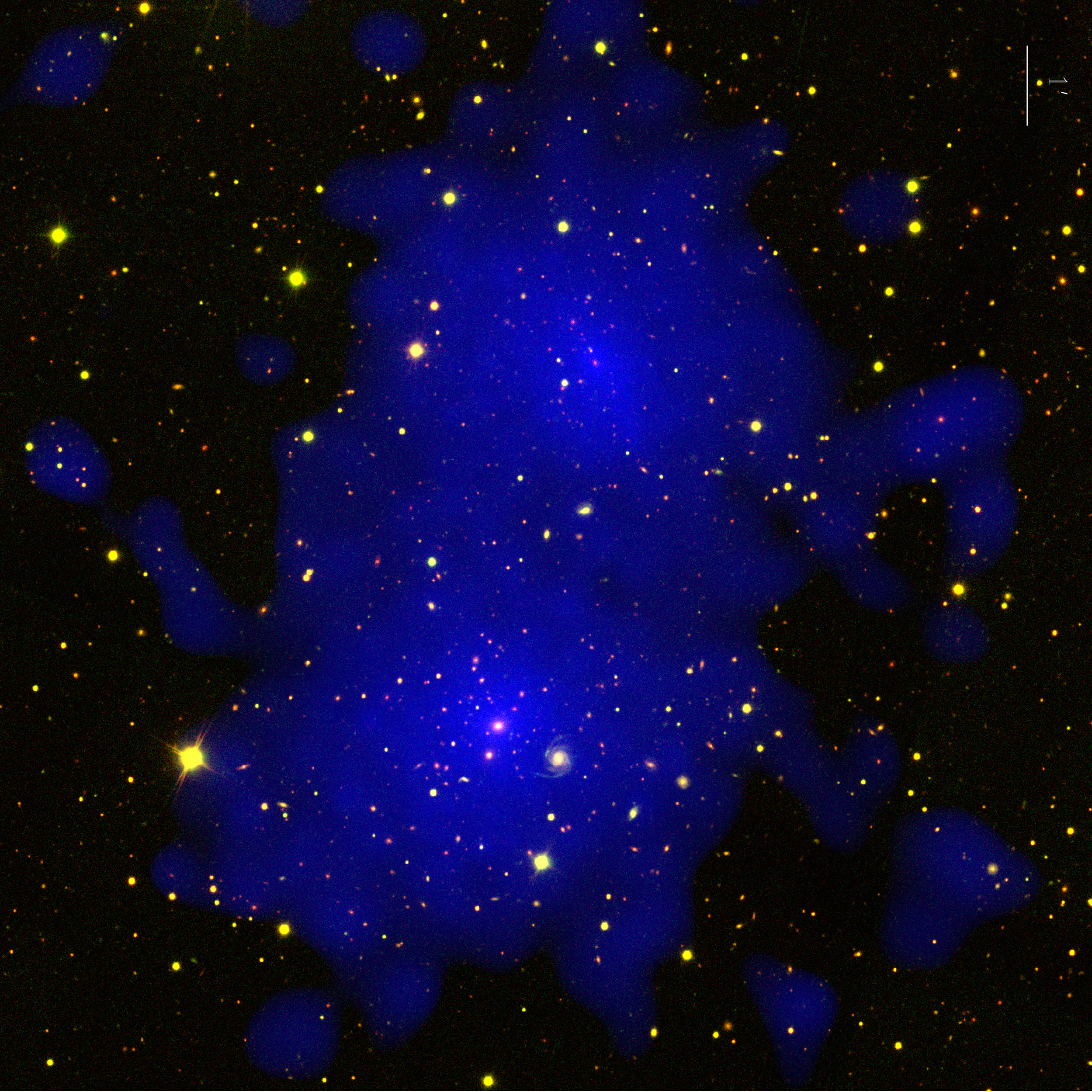}
}
\caption{Multi-wavelength montage of the two cluster field: SDSS r
  (red) and g (green) bands, as well as 0.5-8 keV diffuse X-ray emission (blue). 
}
\label{fig:o_x}
\end{figure*}

We map out the diffuse X-ray emission in the field by exercising 
the sources in the count, background, and exposure maps. We remove a 
region of twice the 70\% energy-encircled radius of the point spread function 
around each source. 
The X-ray intensities in these 
removed regions are interpolated from the surrounding areas in the 
smoothing process of the images. We use 
 \textsc{csmooth} with the Gaussian scale calculated adaptively to achieve
a non-X-ray background-subtracted S/N ratio $\gtrsim 4$ for the count image
in the S (0.5-2 keV) band. The same scale map is used to smooth the count, background, and
exposure maps in the 0.5-1 keV and 1-2 keV bands. These smoothed
images in each band are then used to construct a background-subtracted
intensity map. The resultant maps in the two bands are added to form
the final map in the S band, which is presented in 
Fig.~\ref{fig:field_dif}. 

\begin{figure*}
\unitlength1.0cm
\centerline{
\includegraphics[width=0.9\linewidth,angle=0]{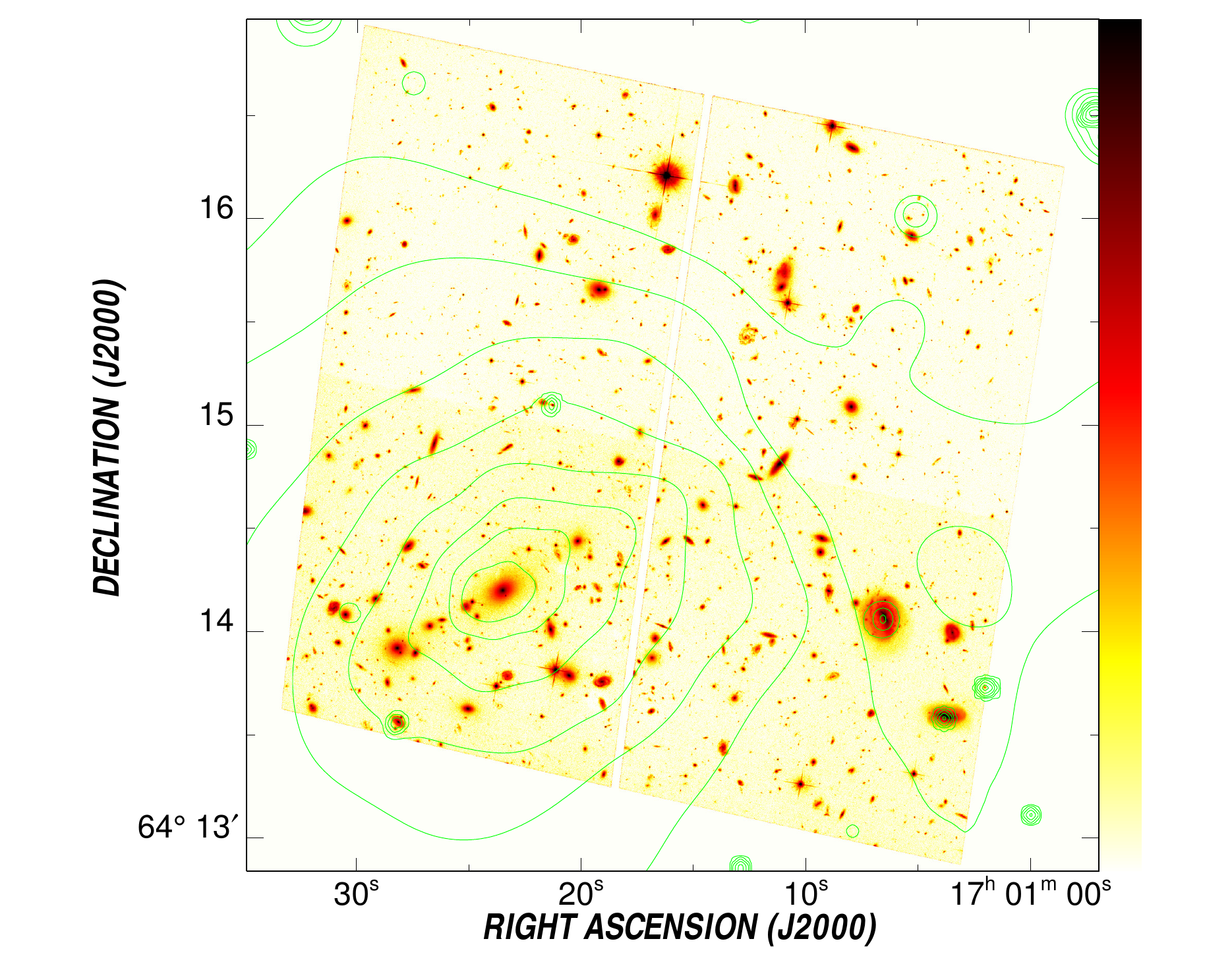}
}
\caption{Close-up of the inner region of Cluster B. Overlaid on the false color
image of the \hst\ optical images (ACS/WFC F814W filter) 
are the intensity contours of the 0.5-8 keV band
emission at 0.5, 1, 2, 4, 8, 16, 32, and 64 above a local background 
of 1.5 (all in units of $10^{-4} {\rm~cts~s^{-1}~arcmin^{-2}}$).
}
\label{fig:o_x_closeup}
\end{figure*}

The X-ray intensity of the clusters peaks at two 
optically red elliptical galaxies (SDSS J170041.75+641258.7 and 
J170123.53+641411.7; Figs.~\ref{fig:o_x} and \ref{fig:o_x_closeup}), 
both of which
are detected as X-ray sources (\#65 or J170041.75+641258.8  and \#126 or J170123.46+641411.9; Table~\ref{t:sou}). 
The X-ray morphology of Cluster A is round and shows no
significant substructure, indicating that the ICM is rather relaxed. By
contrast, the inner region of Cluster B shows a strongly elongated X-ray
morphology, which is lopsided toward to the northwest (Fig.~\ref{fig:o_x}),
as is clearly seen in the close-up presented in 
Fig.~\ref{fig:o_x_closeup}. On the opposite side of this direction
is another giant elliptical (SDSS J170128.23+641354.8). 
Thus this X-ray substructure most likely represents a recent subcluster 
merger. On larger scales ($r
\gtrsim 1^\prime$), the X-ray morphology appears relaxed.

\begin{figure*}
\unitlength1.0cm
\centerline{
\includegraphics[width=0.45\linewidth,angle=0]{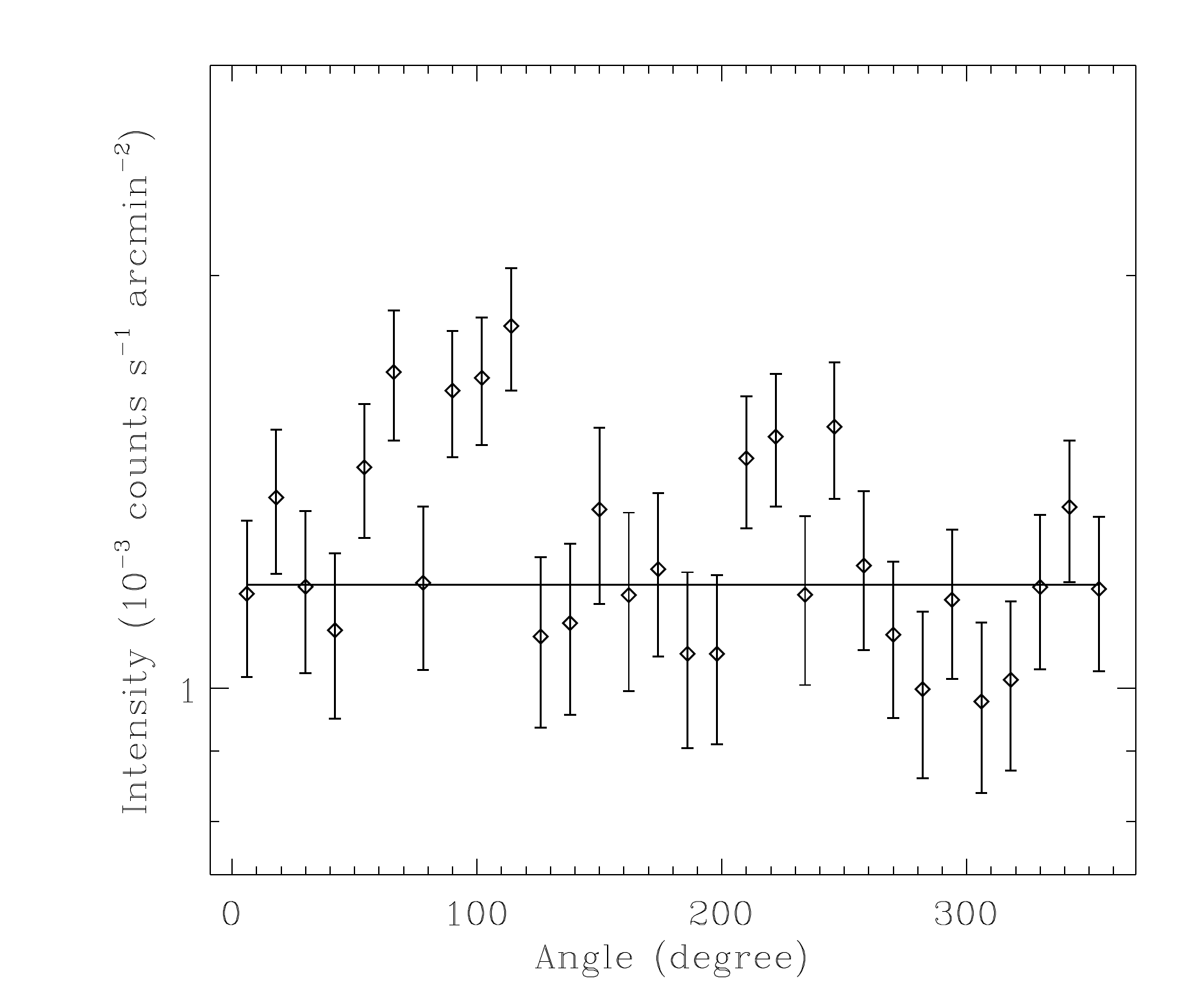}
\includegraphics[width=0.45\linewidth,angle=0]{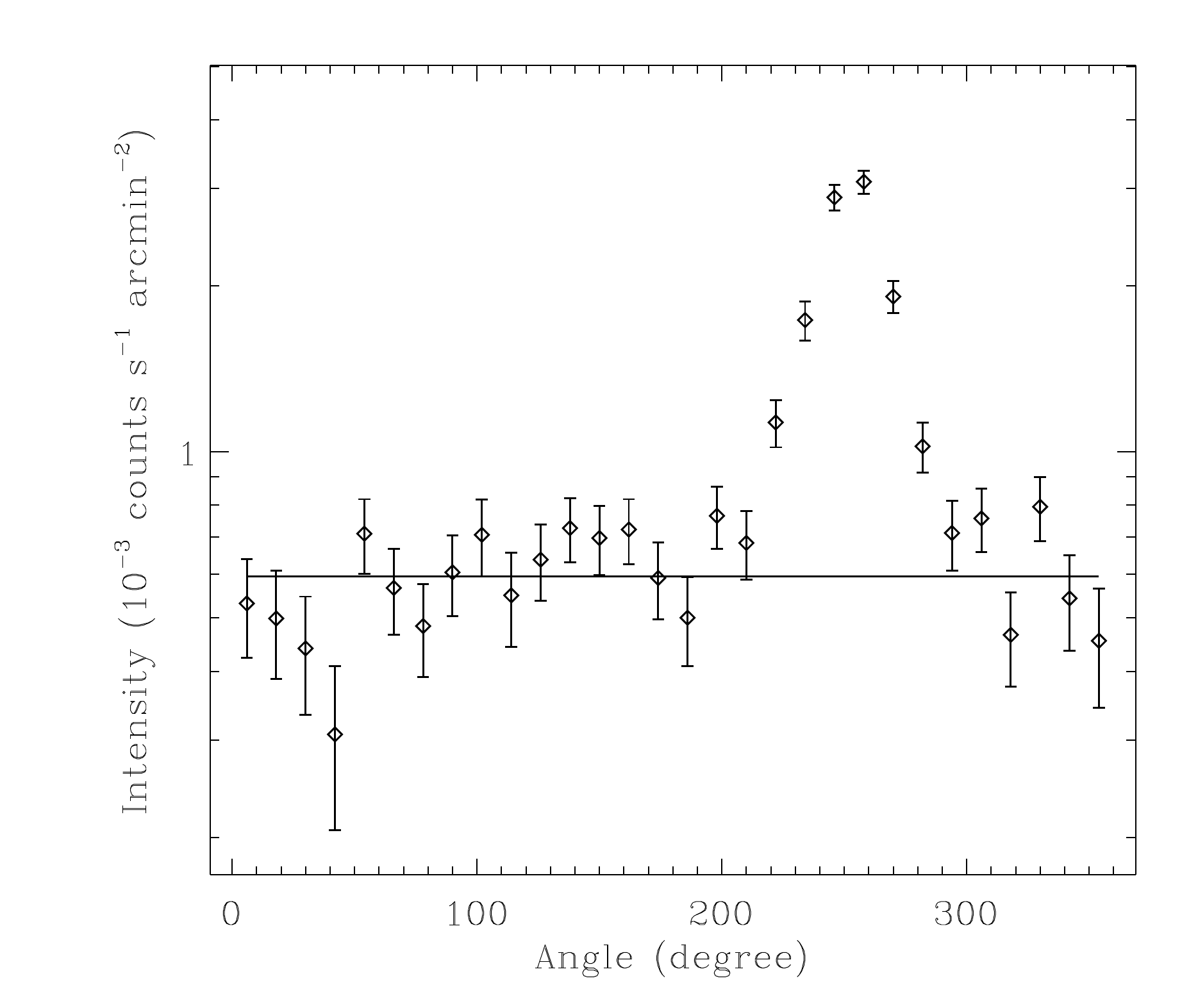}
}
\caption{Azimuthal distributions of the diffuse X-ray intensity of
 Clusters A (left panel) and B (right) in the 0.5-2 keV band.
For each cluster, the data used for the distribution calculation are
extracted from a 1.6$^\prime$-wide annulus with its middle radius
corresponding to the projected off-center distance of \xs;
 the horizontal line
represents the best-fit intensity of the data, excluding the $\pm
45^\circ$ pie around the direction [$75^\circ$ or $255^\circ$ (East from North)]
from the center of the considered cluster (Cluster A or B)
to the center of the other one. Because \xs\ is considerably closer to
Cluster A than to Cluster B, the best-fit average intensity around the former 
is greater than that around the latter. For the same reason, 
the peak (due
to Cluster A) in the right panel is higher than that (due to Cluster B) in the left panel.
}
\label{fig:ad}
\end{figure*}

Clearly, there is a projected overlap between the X-ray
emissions from the two clusters. But this overlap is significant
only in a limited region. As illustrated in
Fig.~\ref{fig:ad}, this region,  viewed from the center of one
cluster, is mostly within $\pm45^\circ$ about the
direction toward the center of the other cluster. 
Excluding this angular interval, the azimuthal
intensity distribution is consistent with a constant (i.e., this null hypothesis
cannot be rejected at $\gtrsim 2\sigma$), at least at the
projected radius of \xs\ in each cluster. At larger radii, the
azimuthal intensity distribution shows a greater dispersion. This may
indicate an increasing clumpiness of the intensity distribution with
the increasing radius. But much of the dispersion  can be
attributed to the cosmic variance of the background/foreground emission (mostly
due to the projected large-scale structure variation from one field to another).
Indeed, the local background fields (Fig.~\ref{fig:field_dif}b) show  comparable intensity
dispersions. Therefore, our detailed analysis in the following is focused on the integrated or
azimuthally-averaged radial properties of the clusters.

\subsection{Spectral analysis of the diffuse X-ray emission}
\label{spectra}
ACIS-I spectra are extracted in concentric annuli, centered
on the peaks of the X-ray emission of each cluster (Fig.~\ref{fig:field_dif}b). The radii
of these annuli are determined so that each contains at least 1000
counts following the non-X-ray background subtraction. When performing the spectral
extraction for each cluster, the overlapping region  with the other
cluster is excluded. The ARFs and RMFs are created for each extracted
spectrum using the CIAO
routines \textsc{mkwarf} and \textsc{mkacisrmf}, allowing us to
account for the variation in the effective area across the chip. 
The spectral analysis uses the software packages \textsc{xspec} (version 12.8).

The local sky background is estimated in the regions chosen to be well
away from the extended emission of the two clusters. After 
the non-X-ray background is subtracted in \textsc{xspec}, 
the ACIS-I spectrum of the local sky X-ray background is modeled as the sum of the
extragalactic contribution and a Galactic foreground. The latter is
characterized with two optically-thin thermal plasma (\textsc{apec}) components: an
unabsorbed one at $kT=0.14$~keV for the Local Hot Bubble and an
absorbed one at $kT=0.6$~keV, representing the Galactic halo
emission. The temperatures of these components are found by fitting
to the {\it ROSAT} all-sky survey spectrum for an annulus around the two clusters (between
0.5-1.0 degrees). 
(The survey data are obtained from the X-ray background tool
at http://heasarc.gsfc.nasa.gov/cgi-bin/Tools/xraybg/xraybg.pl.) The
extragalactic contribution from unresolved point sources is modeled as
a power law of index 1.4. The fit to the ACIS-I 
spectrum then determines the normalizations of the combined thermal
component and the power law. The best-fit model is then
included in the analysis of the non-X-ray-subtracted on-cluster spectra, 
accounting for their geometrical area differences from the local
background region and assuming that the spectral shape and surface 
brightness of the sky X-ray background remain the same across the \chandra\
field of the present study.

This double-subtraction method of the background contributions
accounts for the variation
of the X-ray effective area across the field, as well as the potential change
of the non-X-ray background across the detector. If one simply subtracted the
{\sl observed} spectrum in the background region (near the edge of the
ACIS-I field of view) from the cluster emission regions (which are
closer to the axis of the observations), the background would be underestimated because the effective area is lower off axis.

\begin{figure*}
   \includegraphics[width=0.45\linewidth,angle=0]{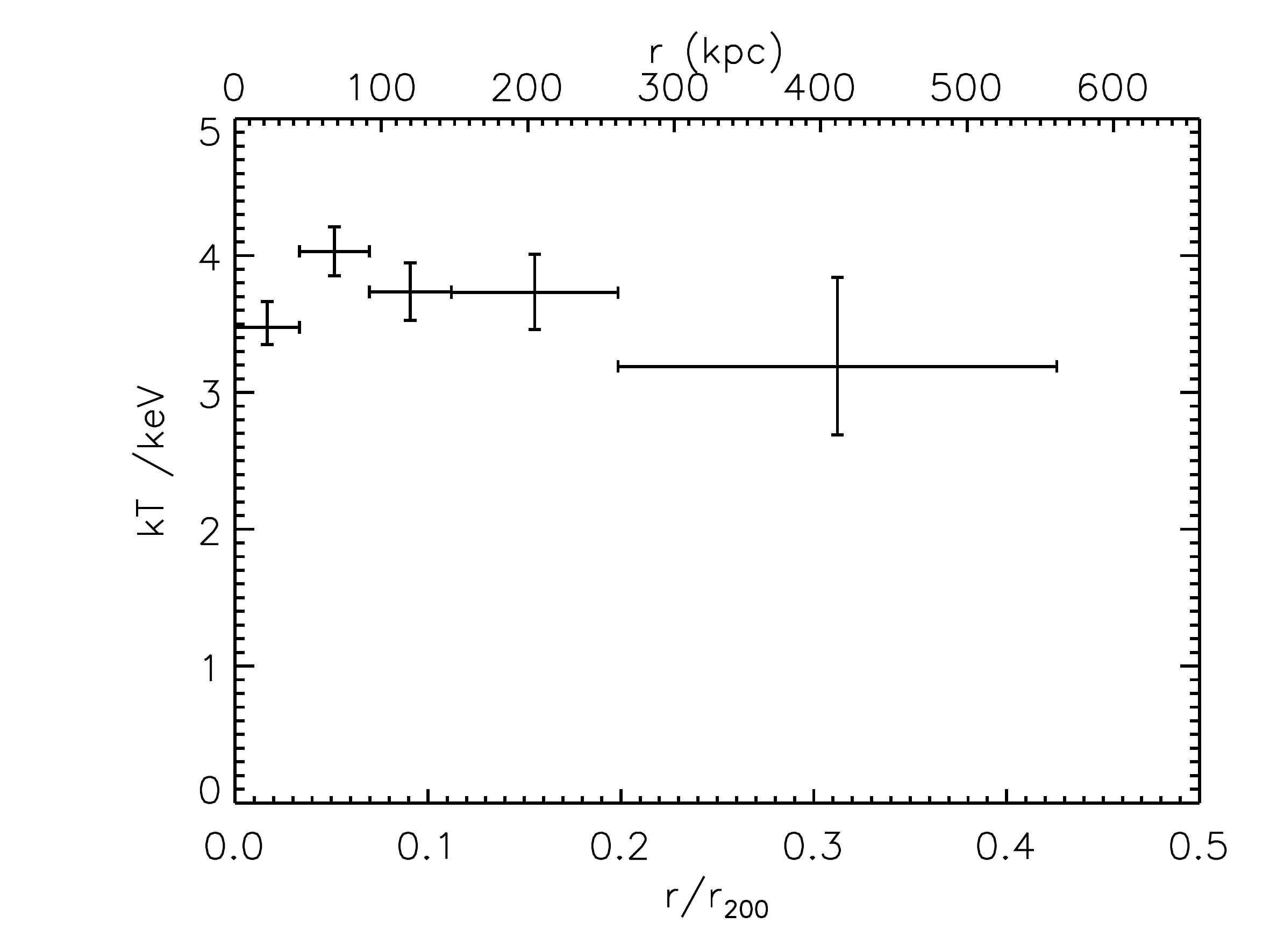}
    \includegraphics[width=0.45\linewidth,angle=0]{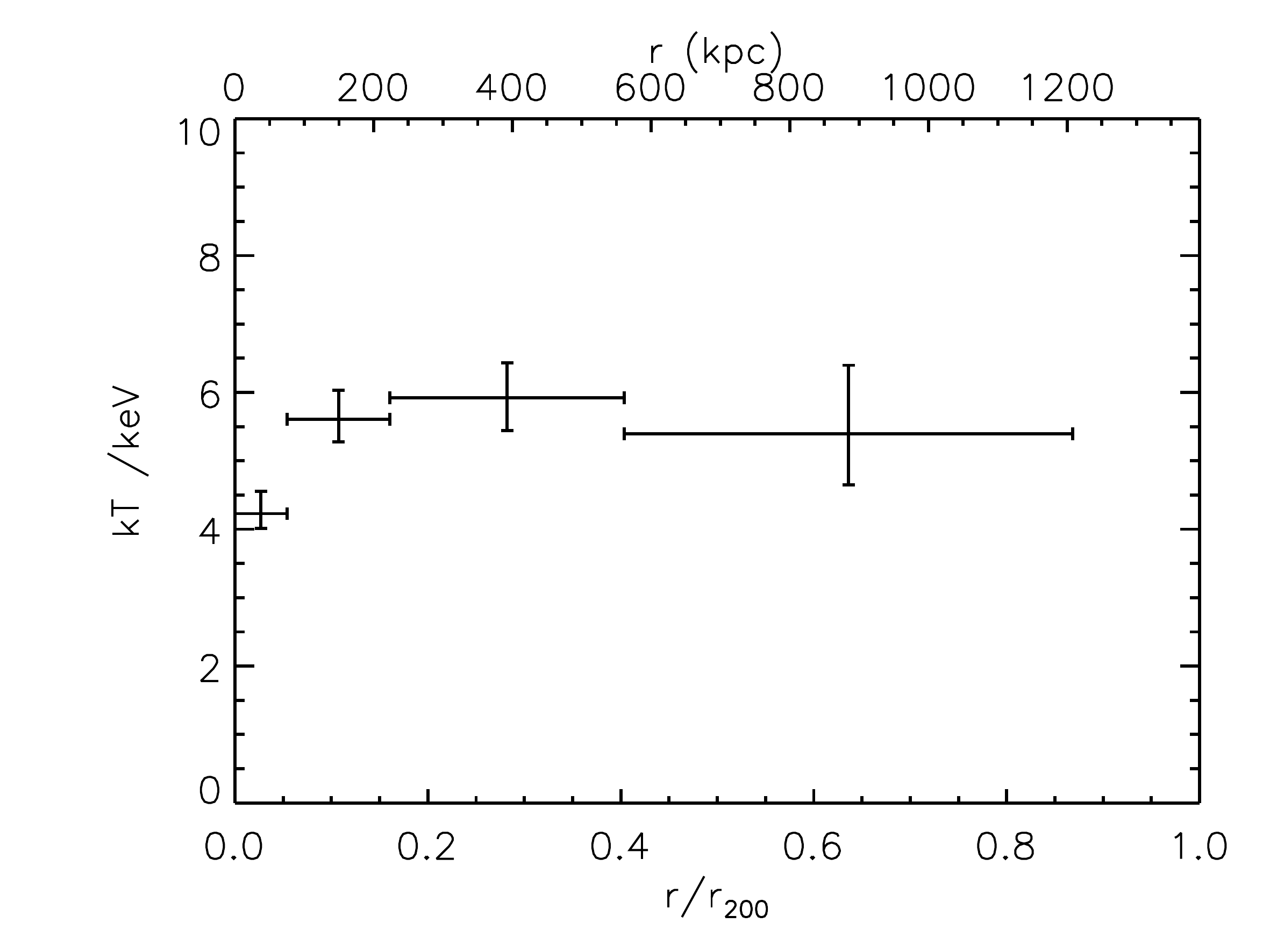}
\includegraphics[width=0.45\linewidth,angle=0]{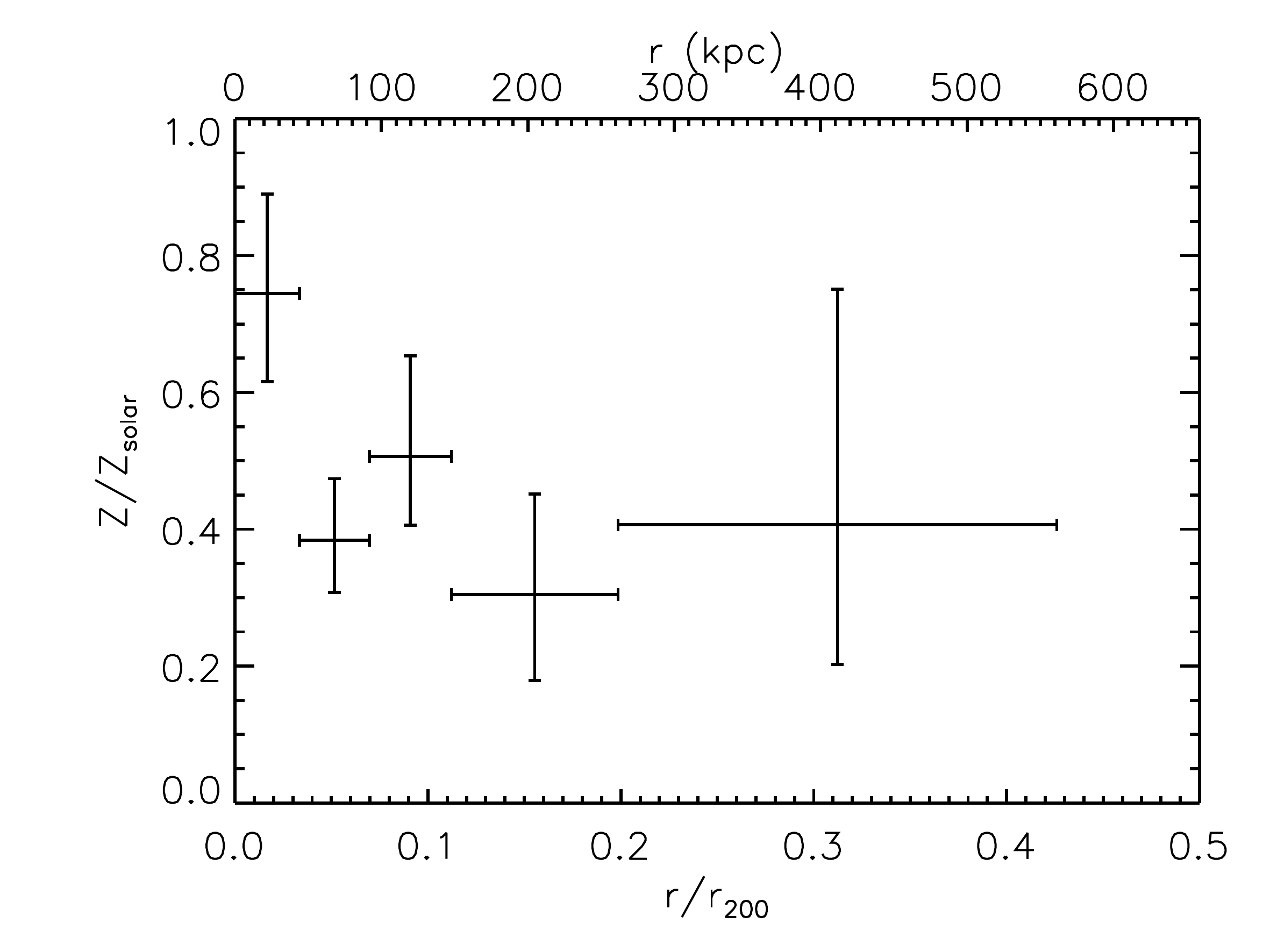} 
\includegraphics[width=0.45\linewidth,angle=0]{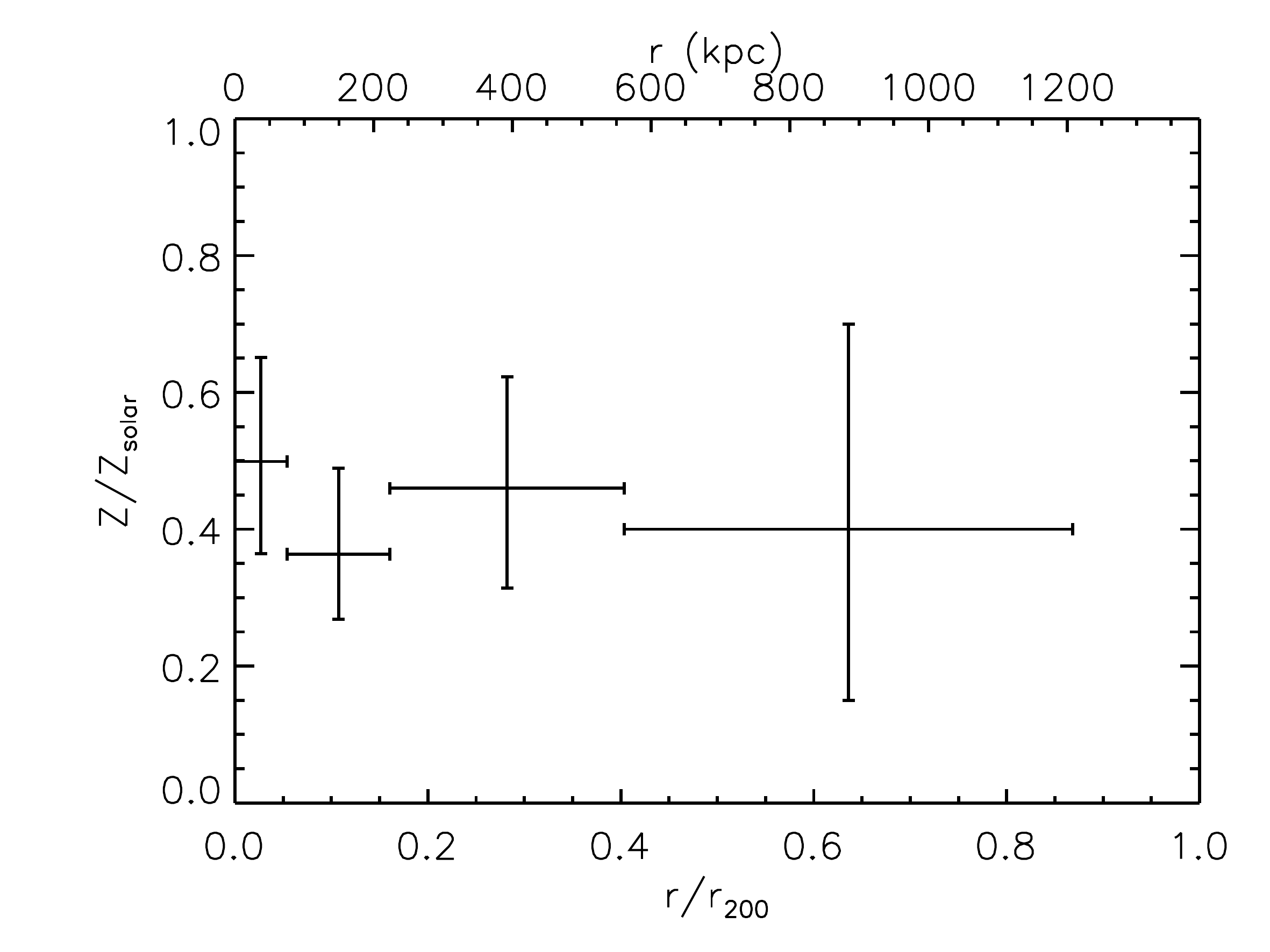}
      \caption{Temperature (upper panels) and metallicity (lower
        panels) for Clusters A (left) and B (right).  }
      \label{TandZandn_profiles}
\end{figure*}

Due to the low counting statistics of the spectral channels, we use the extended C-statistic when
performing the spectral fits. The column density is fixed to the Leiden/Argentine/Bonn
survey value of 2.2$\times$10$^{20}$ cm$^{-2}$~\citep{Kalberla2005}, and the redshift is
also fixed to the value of each cluster. The temperature, metal abundance, and normalization of the
\textsc{apec} component are allowed to be free parameters, and the
profiles of temperature and metallicity are shown in
Fig.~\ref{TandZandn_profiles}. The metal abundance is relative to the solar one 
given by ~\citet{Anders1989}.

The cool cores of the two clusters are resolved. The metallicity
peaks at the cores. With the core bins excluded, we find the average
temperatures as listed in Table~\ref{t:para}, although there is
marginal evidence for declining temperature with increasing radius
for Cluster A. 

From the average temperatures, we estimate $M_{200}$ and
$r_{200}$, using the scaling relations of \citet{Arnaud2005}:
$h(z)M_{200}=5.34~M_\odot (kT/5{\rm~keV})^{1.72 } 10^{14}$, and
$h(z)r_{200}=1674{\rm~kpc} (kT/5{\rm~keV})^{0.57}$. 
Folded through these estimates are the statistical uncertainties
in the temperatures and in the scaling relations
\citep{Arnaud2005}. The results are also included in Table~\ref{t:para}.

\subsection{Projected radial distributions of the diffuse X-ray emission }

\begin{figure*}
\unitlength1.0cm
\centerline{
\includegraphics[width=0.9\linewidth,angle=0]{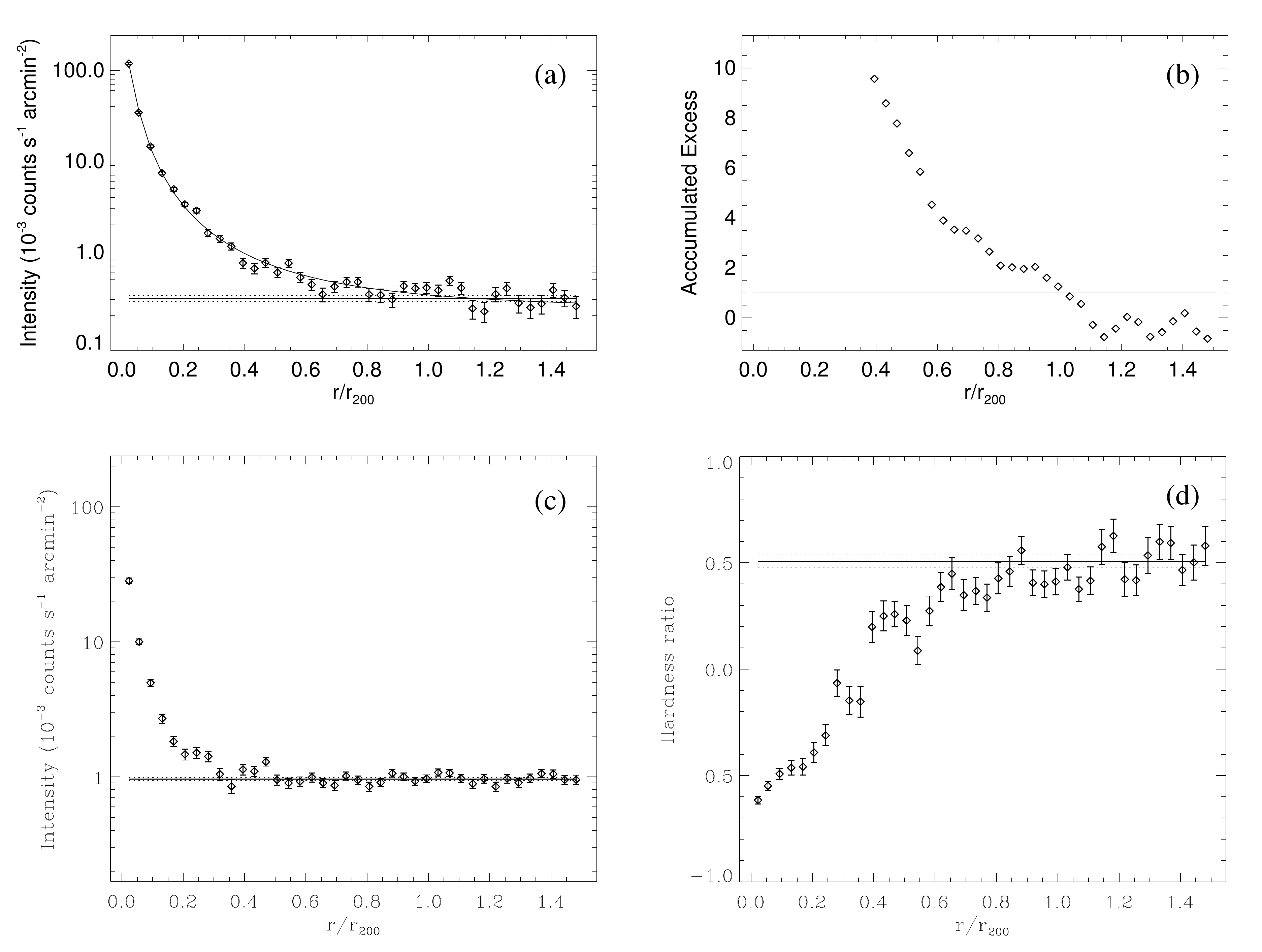}
}
\caption{Projected radial distributions of the diffuse X-ray emission
from Clusters A: the intensity in the S band (a) and its accumulated 
significance (the ratio of the accumulated intensity to its error 
in units of $\sigma$) above the local background; 
the intensity in the H band (c) and the hardness ratio of the 
intensities in the two bands (d). The horizontal solid (dotted) lines 
in the panels (a), (c) and (d) mark
the local background ($\pm 1 \sigma$ uncertainties) estimated
with the data in the $r/r_{200} =1.2-1.5$ range. The two horizontal solid lines
in the panel (b) mark the 1- and 2-$\sigma$ significance levels. The
significance is accumulated inward from $r/r_{200} = 1.5$.
}
\label{fig:rbp_A}
\end{figure*}

\begin{figure*}
\unitlength1.0cm
\centerline{
\includegraphics[width=0.9\linewidth,angle=0]{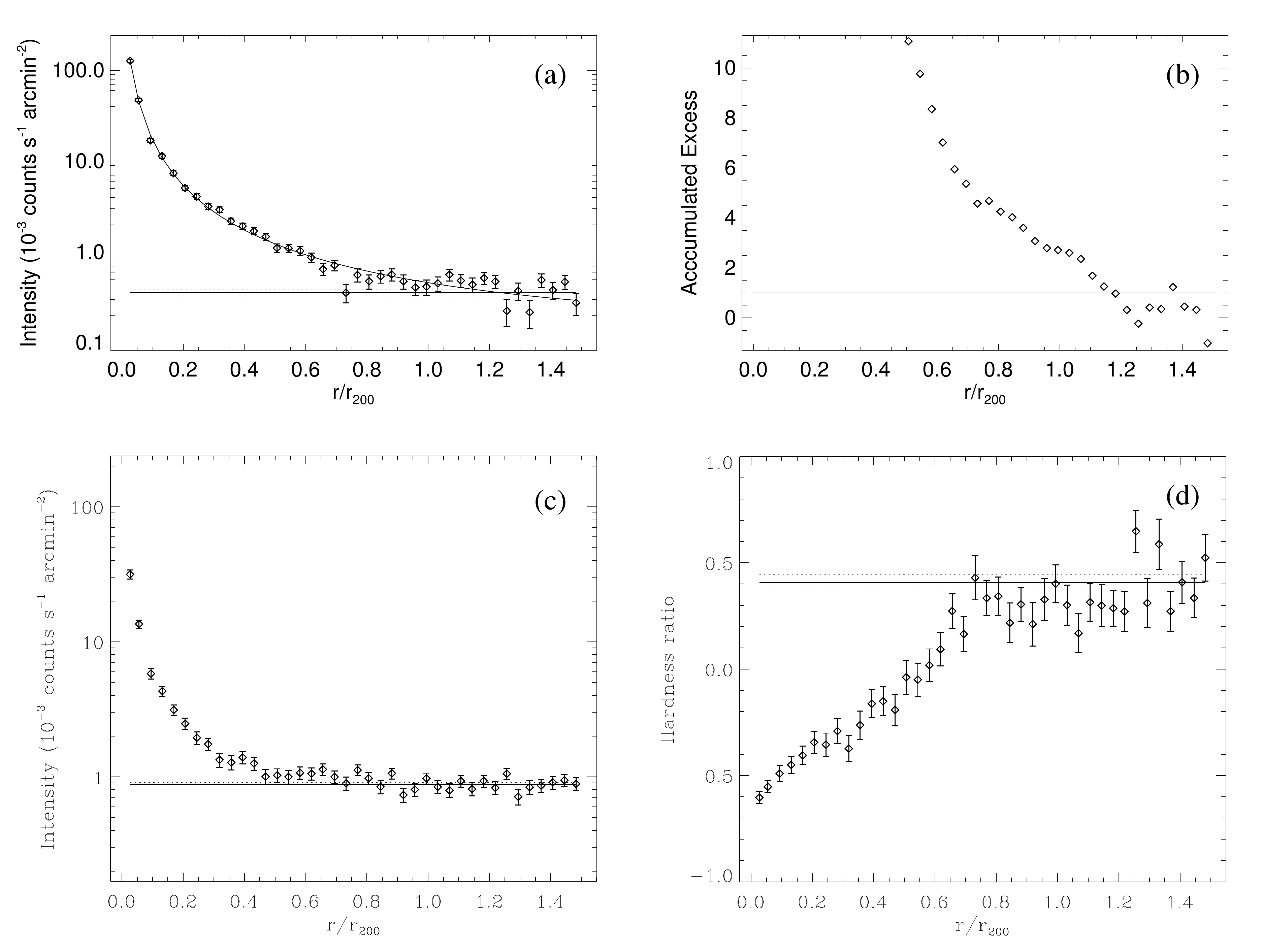}
}
\caption{Same as Fig.~\ref{fig:rbp_A}, but for Clusters B.
}
\label{fig:rbp_B}
\end{figure*}

Figs.~\ref{fig:rbp_A} and ~\ref{fig:rbp_B} present the projected radial 
distributions of the diffuse
X-ray emission around the centers of the two clusters. 
The construction of the distributions for each cluster uses 
only the data in the half of
the circle (with the projected radius $r=1.5r_{200}$) away from the other cluster [e.g.,  between
165$^\circ$ and 345$^\circ$ (East from North) for Cluster A]. The 
distributions appear consistent in the two data
 sets, when viewed separately (\S~\ref{s:obs}). 
We fit the S band intensity profiles [panel (a) in each of the two figures]
with the standard $\beta$-model of the form 
~\citep{Cavaliere1976}:
\begin{equation} 
I = I_{o} \left(1+ {r^2 \over r_c^2}\right)^{1/2-3\beta}.
\label{e:beta}
\end{equation}  
The best-fit parameters are included in Table~\ref{t:para}.
The model typically gives 
a good characterization of a relaxed cluster. But for Cluster A, 
there are clear systematic
deviations of the observed profiles from the model at outer radii
$r/r_{200} \sim 1$. Excluding the six consecutive data points above the
model curve between $r/r_{200} = 0.9-1.1$, the fit improves
significantly ($\chi^2/d.o.f.=59/30$); the fitted model parameters
also change significantly.
Here we have accounted for only statistical counting statistics, not the
discreteness of undetected sources in the field, which is likely  an
important source of uncertainties. 
Indeed, the profiles in the H band [panel (c)] are more
structured, as they are subject to more effects due to the 
discreteness and cosmic variance of sources just below our detection
threshold.
Furthermore, the X-ray intensity in the H band also depends more
sensitively on the ICM temperature (hence its variation) than in the S band. 
Therefore, we do not present the $\beta$ model fitting results for the
H band here.

To assess the radial extent of the cluster emission, we use 
the accumulated significance distribution of the S band emission above 
the local background, which is $\chi^2$-fitted in the $r/r_{200} =1.2-1.5$ 
range for each cluster (Figs.~\ref{fig:rbp_A}b or ~\ref{fig:rbp_B}b). 
The use of this accumulated excess significance distribution minimizes the effect of 
fluctuations, which depend on the binning of the data in constructing an
intensity profile. If the 2$\sigma$ significance of the emission excess
is considered as a threshold, 
we then find that the extents of Clusters A and B are about $ 0.9$ and
$1.1r_{200}$ , respectively (Figs.~\ref{fig:rbp_A}b or ~\ref{fig:rbp_B}b). These are likely conservative estimates of the actual extents. One
indication for this is that the accumulated significance still decreases {\sl
monotonically}
with the increasing radius beyond the above radii, down to 
the $\sim 1\sigma$ significance level (Figs.~\ref{fig:rbp_A}b or ~\ref{fig:rbp_B}b). Our adopted local background may also be 
an over-estimate, because the $r/r_{200} =1.2-1.5$ range could be contaminated
by the cluster emission. Accounting for these factors, the actual
extents of the emission excess are likely to be $\gtrsim  1$ and $1.1r_{200}$
for Clusters A and B, respectively. This conclusion is
consistent with the radial
distributions of the X-ray emission hardness ratio, (H-S)/(H+S), as shown
in Figs.~\ref{fig:rbp_A}d and ~\ref{fig:rbp_B}d. The
softening of the X-ray emission with the decreasing radius is
primarily due to
the increasing contribution from the clusters, relative to the local
sky background. The hardness ratio seems to be systematically below 
the average value of the local background for the entire radial 
range up to $\sim 1.1$ (or 1.2) $r_{200}$ for Cluster A (or B). 


\section{Discussion}

The results presented in the last section represent the direct
measurements or characterizations that we obtain with the \chandra\ data. We now proceed
to extend these results by incorporating
additional assumptions.

\subsection{Radial density profiles of the hot intracluster medium}

\begin{figure*}
\includegraphics[width=0.45\linewidth,angle=0]{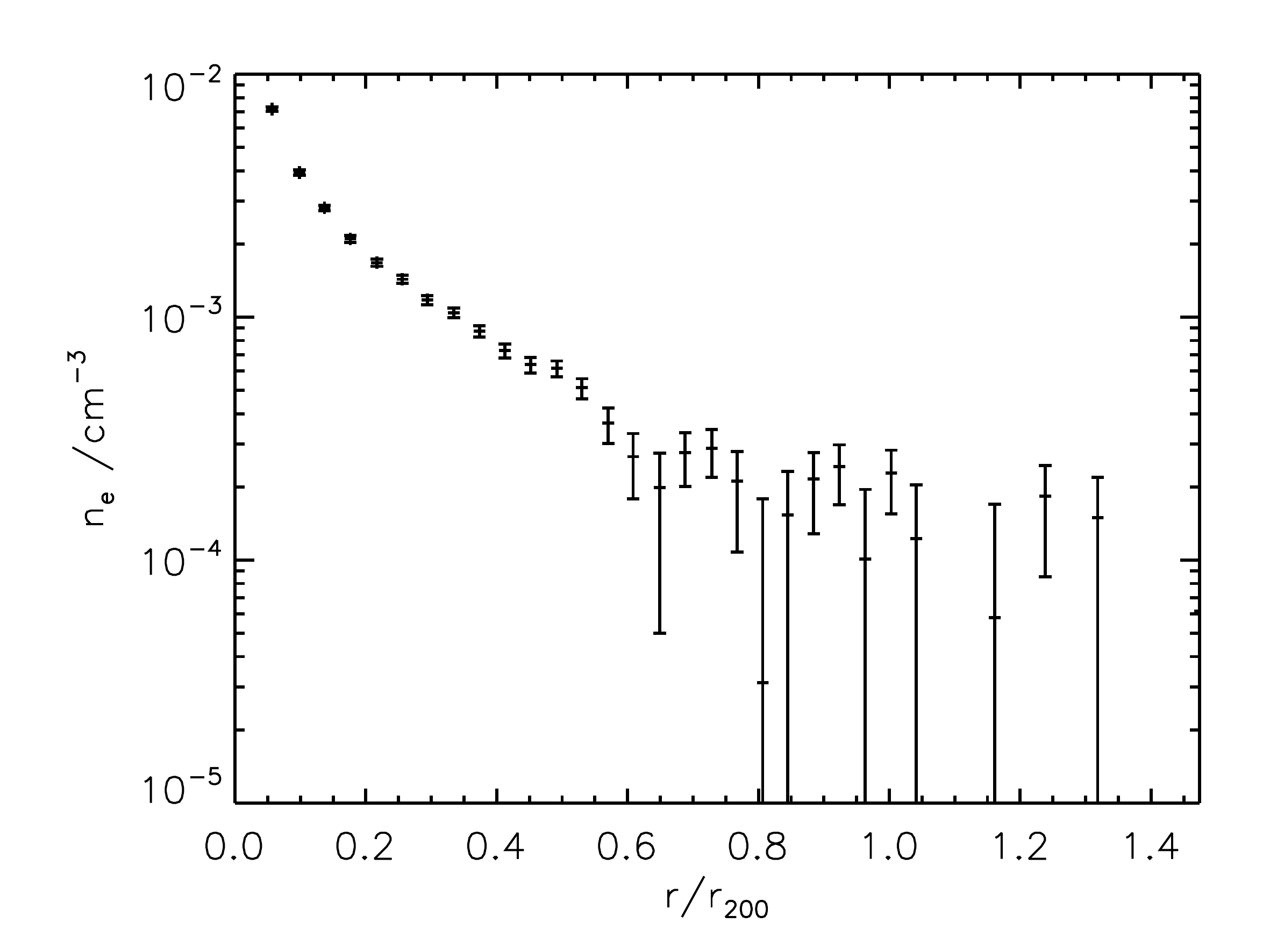}
\includegraphics[width=0.45\linewidth,angle=0]{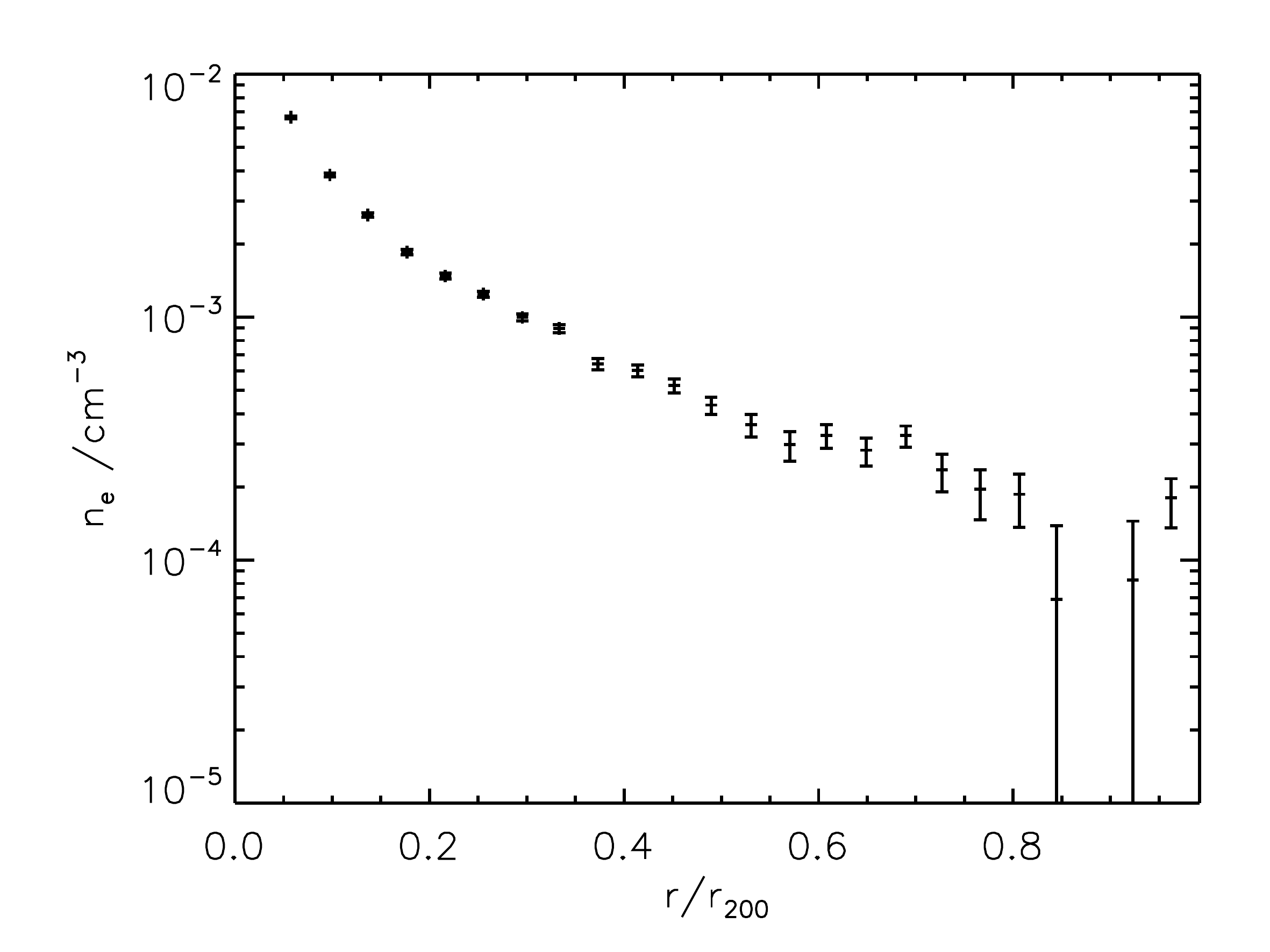} 
      \caption{ De-projected electron density profiles. The rest is
        the same as in Fig.~\ref{TandZandn_profiles}.  }
      \label{density_profiles}
\end{figure*}

We have shown in \S~\ref{s:res} that the morphology of the diffuse
X-ray emission from the clusters are consistent with being azimuthally
symmetric, at least away from the core regions. Thus it may be reasonable
to assume an approximate spherical symmetry to 
infer the de-projected radial profile of the ICM electron density
as a function of the physical radius $r$ for each
cluster. The S band intensity is insensitive to the plasma
temperature (e.g., changing by only about 6 percent for a temperature
range of 1.5-8.0~keV). We can thus first convert the intensity into an
\textsc{apec} emission measure, and then calculate the density
profile. We do this for both clusters in the directions away from the
overlapping region. The resultant density profiles are shown in
Fig.~\ref{density_profiles}. The average background level outside
1.2$r_{200}$ (shown by the horizontal line in Figs.~\ref{fig:rbp_A}
and~\ref{fig:rbp_B}) is
subtracted. 
Table~\ref{t:para} includes
the integrated column densities of the hot ICM along the sightline
toward \xs. 

Based on the spatially resolved temperature and density profiles
(Figs.~\ref{TandZandn_profiles} and~\ref{density_profiles}), we further
conduct the hydrostatic mass analysis. We follow the procedure of
\citet{Schmidt2007}, in which the total mass profile is assumed to be an
NFW profile~\citep{Navarro1997} and the density profile is used to predict
the temperatures in the annuli around each cluster, assuming hydrostatic equilibrium. The
best-fit mass profile corresponds to the predicted temperature profile
which matches the observed one best. The method avoids
the need to use parametric fitting functions for the temperature and
density profiles and the need to extrapolate them out to $r_{200}$, which
could lead to large uncertainties in the priors~\citep{Allen2008}. We find
that both clusters are well fit by the NFW profile, and the resultant
$M_{200}$ and $r_{200}$ values are completely consistent with those
obtained from the scaling relations (\S~3.2).

\subsection{Radial entropy profiles}

The entropy profile of a cluster can provide direct information about the
heating process of the ICM, especially the effect due to galactic
mechanical energy feedbacks. But as shown in \S~\ref{s:res}, we cannot
measure the ICM temperature at the largest radii, where cluster
emission is detected, because of the limited counting statistics. If,
however, we introduce a prior by assuming the universal pressure
profile of \citet{Arnaud2010} to hold, we can then use our density profile to infer the temperature profile and
thus the entropy profile. Note that the clusters studied in
\citet{Arnaud2010} are all at redshift below 0.2, so we only
investigate Cluster A here as the uncertainties in extrapolating out to the much higher redshift of z=0.45 are unknown. 

\begin{figure}
\includegraphics[width=\linewidth,angle=0]{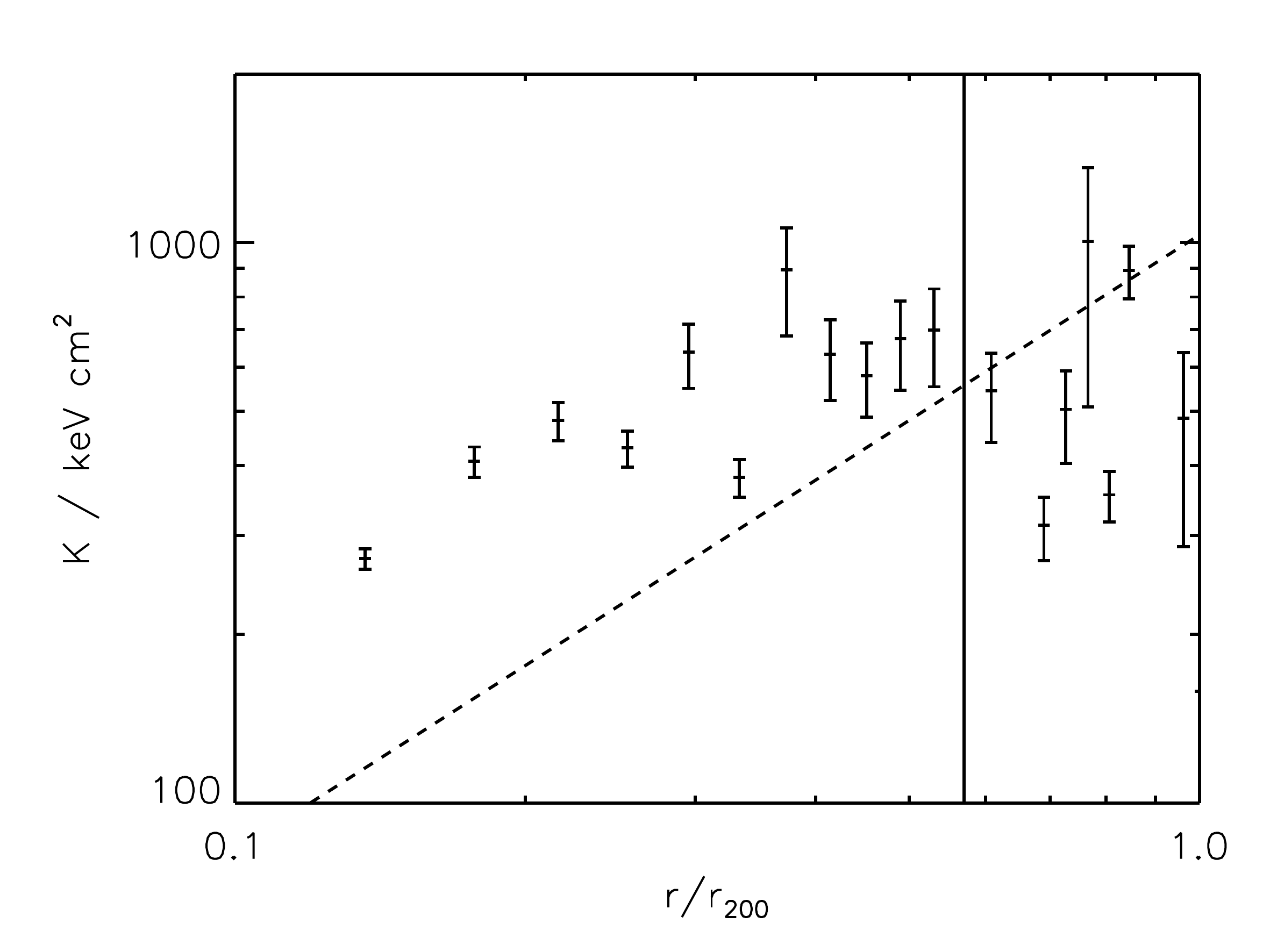}
\caption{Entropy profile for Cluster A, derived from the de-projected density profile in Fig. \ref{density_profiles} and the universal pressure profile of \citet{Arnaud2010}. The dashed line shows the theoretical power law relation for purely gravitational hierarchical structure formation from~\citet{Voit2005}.}
      \label{entropy_profile}
\end{figure}

The de-projected temperature profile, $kT = P/n_{e}$, can be inferred from the universal pressure profile and the
de-projected density profile that we have measured,
while the de-projected entropy profile can be estimated from
\begin{equation}
K = P/n_{e}^{5/3}.
\end{equation}
Fig. \ref{entropy_profile} shows the inferred  entropy profile for the
cluster. Note that the error bars are derived from the
errors in the density profile, and are underestimates because the
uncertainty in the pressure profile is not accounted for. We compare
this crude entropy profile to the power law prediction of purely
gravitational hierarchical structure formation from \citet{Voit2005}.
We see that there is a central excess in entropy within 0.5$r_{200}$,
as has been found for samples of nearby clusters using {\sl XMM-Newton}
\citep{Pratt2010} and {\sl Suzaku} observations \citep{Walker2013}. 
\citet{Eckert2013} have also found the same central entropy excess by combining stacked \ROSAT\ density profile with stacked Planck pressure profiles. 
Outside of 0.5$r_{200}$ the scatter is large due to the uncertainties in the
density de-projection. Nevertheless, the points seem to lie systematically
below the baseline entropy profile. These findings suggest that the
non-gravitational feedback must be important in the 
inner regions, while the ICM in outer
  regions near the virial radii may not be hydrostatic and/or
  ionization equilibrium and may be rather clumpy and multi-phased.

\section{Summary}
We have conducted a detailed analysis of the \chandra\ observations of 
the two foreground clusters of \xs: Abell 2246 (Cluster A) and GMBCG
J255.34805+64.23661 (B). Table~\ref{t:para} includes the key parameters of the
clusters, measured or inferred from the present work.  
We summarize our main results and conclusions in the following:

\begin{itemize}
\item Spatially, the two clusters are well resolved in the observations. In
  particular, the X-ray emission is found to be centered on a red
  early-type galaxy in each of the clusters. Whereas Cluster A appears well relaxed
  morphologically, the core of Cluster B is strongly elongated,
  probably as a result of a recent subcluster merger. The X-ray
  morphology of the two clusters at the impact distances of \xs\ is consistent with being azimuthally symmetric.

\item Spectrally, a cool core is detected in each of the two
  clusters. Cluster A also shows a significant abundance
  enhancement in the core.
  Outside the cores, the ICM has a mean metal abundance of $\sim 0.4$ solar
  and a mean temperature of $\sim 4.0$~keV for 
  Cluster A (or 5.5~keV for Cluster B). We have estimated the gravitational
  masses and the virial radii of the clusters from the ICM temperatures. 

\item Each of the two clusters shows significant X-ray emission
  extending to about $r \sim  r_{200}$. Assuming a spherical
  symmetry  of the ICM, we have obtained its radial density and temperature distributions for each cluster. By adopting a ``universal''
  pressure profile, we have further shown that the radial entropy
  distribution of Cluster A is
  substantially flatter than what might be expected from purely
  gravitational hierarchical structure formation, suggesting the importance
of galactic feedback in cluster inner regions and non-equilibrium of the 
ICM in outer regions. 
\end{itemize}

The above results and conclusions represent an essential step to characterize
the thermal, chemical, and dynamic states of the clusters. In
particular, we can now estimate the density and temperature properties of
the hot ICM along the sightline of \xs\ through the two clusters at
the impact distances of 0.36 and 0.8$r_{200}$.

With the upcoming \HST/COS observations along the
\xs\ sightline, we will be able to cross-correlate the
  properties determined from the UV absorption lines with the 
X-ray properties of the clusters.
The \HST/COS absorption line spectroscopy is well-suited for investigating 
the warm phase of the ICM, reaching column density limits of 
$\sim10^{12} {\rm~cm^{-2}}$ for HI and $\sim10^{13.5} {\rm~cm^{-2}}$ for OVI 
and to simultaneously explore other multiple transitions from a wide range 
of atoms/ions. The OVI doublet, in particular,  
is sensitive to the so-called transition temperature gas at $\sim 10^{5}$ K
and is hence ideal
for studying the interplay between warm and hot gas components. The combination of the lines, in terms of
their column densities and velocity centroids/widths, can give powerful diagnostics of the absorbers and their relationship to the surrounding hot ICM~\citep[e.g.,][]{Tripp2008}.
Therefore, we will be able to examine the thermal, chemical,
ionization, and kinematic properties as well as the baryon content of
the multiphase ICM. Two additional pairs of FUV-bright QSO/clusters
pairs at similar redshifts have been proved for COS observations.
These studies will nicely complement the existing {\sl HST}/COS observations
of the very nearby Virgo cluster, although they do not allow for 
access to the OVI doublet \citep{Yoon2012}. 
25 Ly-$\alpha$ absorbers (N$_{HI} \gtrsim 10^{13.1-15.4} {\rm~cm^{-2}}$) 
are detected toward 9 QSO sight-lines. It is concluded that the warm gas covering fraction ($\sim 
100\%$ for N$_{HI} = 10^{13.1} {\rm~cm^{-2}}$) is in agreement with cosmological simulations.
Comparisons of the results will then provide
statistical insights into the ICM/CGM properties and their dependence on various cluster
properties and impact parameters.

\section*{Acknowledgments}
We thank the anonymous referee for valuable comments that led to an
improved presentation of the paper. QDW thanks the Institute of Astronomy for the hospitality 
and the award of a Raymond and Beverley Sackler
Distinguished Visitor fellowship. The research is also partly
supported by NASA and CXC via the grants NNX12AI48G and TM3-14006X.

\appendix

\section{Discrete X-ray Source Detection}
\label{a:app}
We search for X-ray sources in the S, H, and B broad bands. 
A combination of source detection 
algorithms are applied: wavelet, sliding-box, and maximum likelihood 
centroid fitting~\citep{Wang2004}. Our final source list contains sources
with local false detection probability less than $10^{-6}$ (due to the 
Poisson fluctuation above the local sky background). 
The source detection, though optimized for point-like sources, 
may include strong peaks of diffuse X-ray emission, 
chiefly associated with the centers of the clusters. 

Table~\ref{t:sou} lists our detected sources. Listed parameters of
the sources are defined in the note to the table. The conversion from a count rate to an
unabsorbed energy flux depends on the source spectrum and foreground
absorption. A characteristic value of the conversion 
is $8 \times 10^{-12}$ 
${\rm~(erg~cm^{-2}~s^{-1}})/({\rm counts~s^{-1}})$ in the B band 
for a power law spectrum of photon index 2 and an absorbing-gas 
column density $N_H \sim 1 \times 10^{21} 
{\rm~cm^{-2}}$ (assuming the solar abundances). This conversion should be 
a good approximation (within a factor of 2) for 
 $\lesssim 3 \times 10^{21} {\rm~cm^{-2}}$. 

\begin{figure}
\centerline{ 
\includegraphics[width=0.95\linewidth,angle=90]{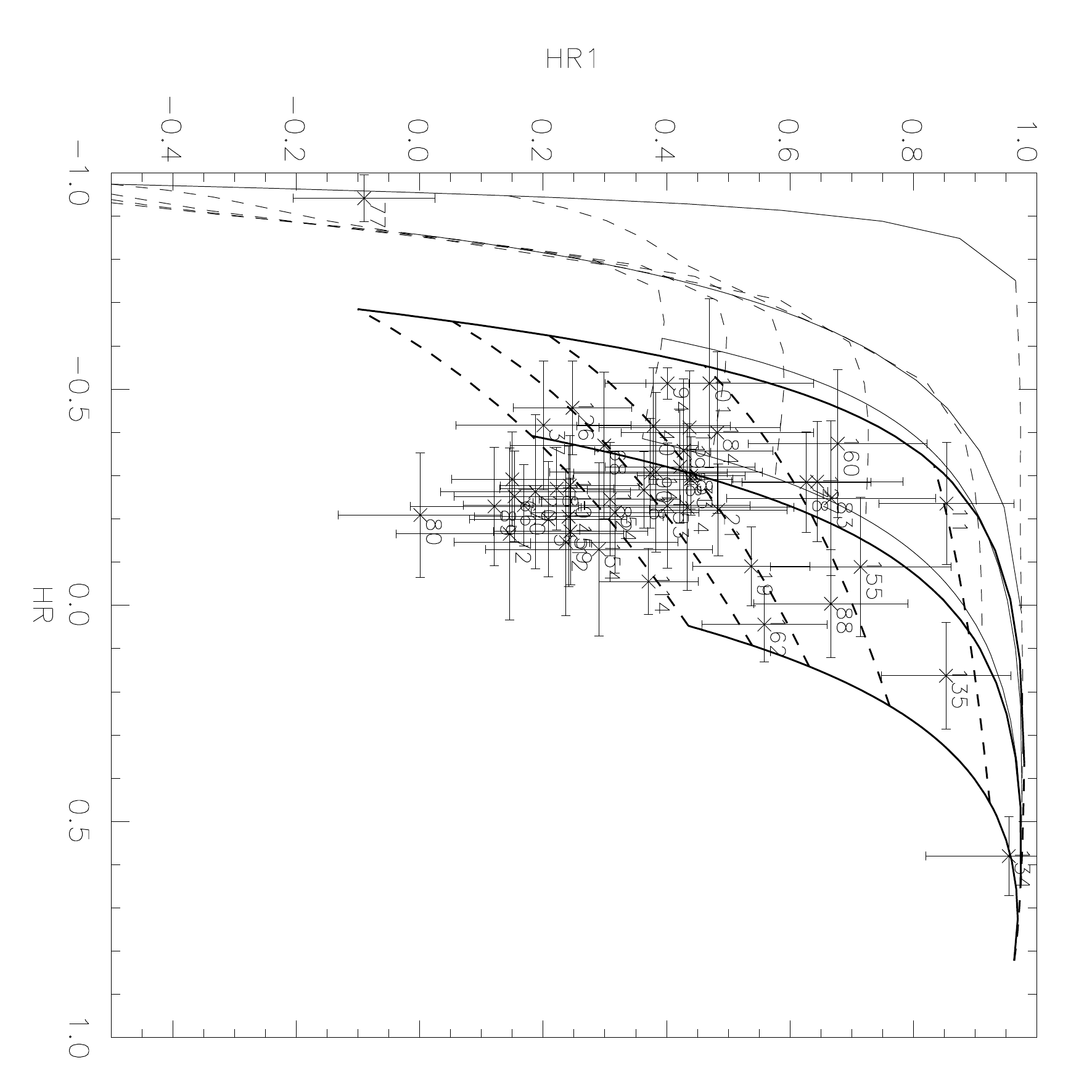}}
\caption{Color-color diagram of X-ray sources with their generic 
numbers (Table~\ref{t:sou}) labeled. The hardness ratios (HR and HR1) 
are defined in the note
to Table~\ref{t:sou}, and the error bars represent 1$\sigma$ uncertainties. 
Also included in the plot are
hardness-ratio models: the solid thick curves are for the power-law model
with photon index equal to 3, 2, and 1, whereas the 
solid thin curves are for the thermal plasma with a temperature
equal to 0.3, 1, 2, and 4 keV, from left to right. The absorbing
gas column densities are 1, 10, 20, 40, 100, and 300
$\times 10^{20} {\rm~cm^{-2}}$ (dashed curves from bottom to top)}
\label{f:figa1}
\end{figure}

The hardness ratios,
compared with spectral models (Fig.~\ref{f:figa1}), may be used to characterize the X-ray 
spectral properties of the sources. Of course, the spectrum of a source can be more complicated and cannot be 
adequately characterized by a simple power law or an optical thermal plasma model
as plotted in Fig.~\ref{f:figa1}; for example, a two-component model consisting of 
a power law plus a thermal plasma may explain those sources located below the
model curves in the HR$=-0.4$ to $0.0$ range. In addition, the spectrum
of a source could also change substantially from one observation to another; 
a good example is \xs\ (\#94 in Table~\ref{t:sou}), the intrinsic absorption of which varies
on order of $10^{22} {\rm~cm^{-2}}$~\citep{Lanzuisi2012}.

\begin{table*}
\begin{center}
\begin{minipage}[t]{4.7in}
\caption{{\sl Chandra} Source List}
\begin{tabular}{lcccrrr} 
\noalign{\smallskip}\noalign{\smallskip}
\hline\hline
\noalign{\smallskip}
\footnotesize  
Source &
CXOU Name &
$\delta_x$ ($''$) &
CR $({\rm cts~ks}^{-1})$ &
HR &
HR1 &
Flag \\
(1) & (2) & (3) & (4) & (5) & (6) & (7) \\
\hline                               
 1 &  J165916.79+640629.9 &  3.5 &$     0.76  \pm   0.14$& --& $ 0.10\pm0.19$ & S \\
   2 &  J165918.70+640715.0 &  3.8 &$     0.38  \pm   0.11$& --& --& S \\
   3 &  J165928.85+640551.1 &  2.9 &$     1.64  \pm   0.16$& $-0.20\pm0.13$ & $ 0.21\pm0.13$ & B \\
   4 &  J165930.94+640631.9 &  2.7 &$     0.85  \pm   0.12$& $-0.23\pm0.19$ & $ 0.43\pm0.17$ & B \\
   5 &  J165931.04+640816.2 &  3.2 &$     0.12  \pm   0.07$& --& --& S \\
   6 &  J165934.84+640955.5 &  2.3 &$     0.18  \pm   0.06$& --& --& B \\
   7 &  J165939.79+640805.0 &  2.5 &$     0.21  \pm   0.06$& --& --& B \\
   8 &  J165939.93+640642.9 &  3.1 &$     0.19  \pm   0.07$& --& --& S \\
   9 &  J165941.23+640533.6 &  2.7 &$     0.30  \pm   0.10$& --& --& S \\
  10 &  J165942.46+640728.6 &  2.2 &$     0.35  \pm   0.08$& --& --& B \\

\multicolumn{7}{c}{...}\\
\multicolumn{7}{c}{Complete table published online}\\
\hline
\end{tabular}

The definition of the bands:
S=S1+S2, H=H1+H2, and B=S+H.
 Column (1): Generic source number. (2): 
{\sl Chandra} X-ray Observatory (unregistered) source name, following the
{\sl Chandra} naming convention and the IAU Recommendation for Nomenclature
(e.g., http://cdsweb.u-strasbg.fr/iau-spec.html). (3): Position 
uncertainty (1$\sigma$) calculated from the maximum likelihood centroiding and an approximate off-axis angle ($r$)
dependent systematic error $0\farcs2+1\farcs4(r/8^\prime)^2$
(an approximation to Fig.~4 of \citet{Feigelson2002}), which are added in
quadrature.  
(4): On-axis source broad-band count rate --- the sum of the 
exposure-corrected count rates in the four
bands and are estimated within (and corrected for) the 70\% 
energy-encircled radius of the point spread function, 
depending on the off-axis angle. 
(5-6): The hardness ratios defined as 
${\rm HR}=({\rm H-S2})/({\rm H+S2})$, and ${\rm HR1}=({\rm S2-S1})/{\rm S}$, 
listed only for values with uncertainties less than 0.2.
(7): The label ``B'', ``S'', or ``H'' marks the band in 
which a source is detected with the most accurate position that is adopted in
Column (2). 
\label{t:sou}
\end{minipage}
\end{center}
\end{table*}

\clearpage

\clearpage

\end{document}